\documentclass[conference]{IEEEconf}

\input epsf
\usepackage{graphicx}
\usepackage{multirow}
\usepackage{titlesec}
\usepackage{balance} 
\usepackage{stfloats}
\usepackage{xcolor}
\usepackage{hyperref}
\usepackage{amsmath}
\usepackage{pgfplots}
\pgfplotsset{compat=1.18}
\usepackage{tikz}
\usepackage{enumitem}
\usetikzlibrary{arrows.meta, positioning}

\renewcommand\thesection{\arabic{section}} 

\renewcommand\thesubsectiondis{\thesection.\arabic{subsection}}

\renewcommand\thesubsubsectiondis{\thesubsectiondis.\arabic{subsubsection}}

\begin{document}

\title{\textbf{\Large Integrating AI into Requirements Quality Learning in Software Engineering Education: A TPACK-Guided Empirical Study\\}}

\author{Hansika Ekanayake Mudiyanselage, Rohan Jai Dharmaraj, Malik Abdul Sami, and Zheying Zhang$^{*}$\\
	\normalsize Software Engineering Research Center (TASE), Tampere University, Finland\\
 	\normalsize {\{hansika.ekanayakemudiyanselage, rohanjai.dharmaraj, malik.sami, zheying.zhang\}}@tuni.fi\\
 	\normalsize *corresponding author
 }


\maketitle
\begin{abstract}
The rapid adoption of generative Artificial Intelligence (AI) in software engineering (SE) practice creates an urgent need for pedagogically grounded approaches to AI integration in SE education, particularly in conceptually intensive subjects such as requirements engineering (RE). While prior studies report students’ perceptions and concerns regarding AI use, systematic investigations grounded in established educational theory remain limited. This study addresses this gap by examining how a multi-agent AI tool can be integrated into an RE course assignment using the Technological Pedagogical Content Knowledge (TPACK) framework as both a design principle and an analytical lens.

Using a mixed-methods design (N=100; 72 submissions analysed), we analyze 
(i) how structured assignment design shapes students’ AI use, (ii) how AI-supported assignment affects students' understanding and application of user story quality criteria, and (iii) how students perceive the benefits and limitations of AI use in the course assignment.


The results show that students engaged selectively with the AI tool and predominantly used it as a support for analysis and evaluation rather than as an automation mechanism. Learning gains were most evident in structurally concrete quality dimensions such as value articulation and testability, while attributes such as negotiability showed mixed effects. Analysis through a TPACK lens revealed that structured pedagogical scaffolding mediated technological affordances, guiding students to engage with core content knowledge rather than relying on automation. Students reported conditional trust, active refinement, and enhanced awareness of quality criteria, alongside moderate usability challenges.

This study provides empirically grounded design guidance for responsible AI integration in RE education and demonstrates how TPACK-guided assignment design can align technological affordances with pedagogical intent and established requirements quality criteria.

\end{abstract}
\IEEEoverridecommandlockouts
\vspace{1.5ex}
\begin{keywords}
\itshape requirements engineering education; AI tool; user story quality; TPACK; INVEST framework
\end{keywords}

%
\IEEEpeerreviewmaketitle

\section{Introduction}

Generative artificial intelligence (AI), particularly large language models (LLMs), is rapidly transforming software engineering (SE) practice. Requirements engineering (RE) is especially 
affected, as its core activities, such as elicitation, analysis, specification, and validation, rely heavily on natural language articulation and interpretation.  
Recent research demonstrates the expanding use of LLMs to generate, refine, and even prioritize user stories and other requirements artifacts \cite{cheng2025generative}\cite{zhang2024llm}\cite{sami2024early}.
While such advances offer opportunities for productivity and support, they also highlight unresolved challenges related to trust, human–AI collaboration, and contextual alignment \cite{parra2025towards}\cite{sami2026bridging}. These developments signal not only a technological shift in RE practice but also a pedagogical challenge for RE education (REE).

Early empirical studies \cite{Guardado_RE_2025} report that students perceive AI tools as helpful for comprehension and productivity, yet concerns remain regarding over-reliance, superficial adoption, and limited critical evaluation of AI-generated outputs. Broader reviews of AI integration in SE and engineering education indicate that many implementations remain exploratory and lack systematic pedagogical grounding and rigorous empirical validation \cite{Sah_CSEET_2024} \cite{Filippi_Motyl_2024}. There is still limited evidence on how AI can be deliberately integrated into conceptually intensive subjects such as RE in ways that preserve and strengthen core analytical competencies.

The challenge, therefore, is not whether AI can support RE tasks, but how it should be pedagogically orchestrated. Effective integration requires alignment between technological affordances, disciplinary knowledge, and instructional design. To address this challenge, we adopt the Technological Pedagogical Content Knowledge (TPACK) framework \cite{koehler2009technological} as both a design principle and an analytical lens. TPACK conceptualizes effective technology integration as the alignment of content knowledge (CK), pedagogical knowledge (PK), and technological knowledge (TK). In the context of REE, CK corresponds to principles, techniques, and practices for RE, including requirements quality frameworks such as INVEST \cite{INVEST}, ISO/IEC/IEEE 29148 \cite{6146379} etc., PK involves structured learning mechanisms such as contrastive analysis and peer review, and TK refers to AI tools supporting RE activities. Only through deliberate alignment of these dimensions can AI function as a scaffold for analytical reasoning rather than as an automation substitute.

This study reports on a TPACK guided integration of a multi-agent AI tool into an assignment in a master-level RE course. The assignment was deliberately designed to sequence manual reasoning before AI use, require comparison between students- and AI-generated requirements, and include iterative refinement, peer review, and structured reflection. Rather than evaluating the AI tool in isolation, we investigate how pedagogical sequencing shapes students’ enactment of AI, how such integration influences their application of requirements quality criteria, and how students critically evaluate AI-generated outputs. Specifically, we address the following research questions:

\begin{description}
    \item [RQ1] How can an AI tool be effectively integrated into course assignments to support structured analysis, reflection, and iterative refinement in students' learning?  
    \item [RQ2] How does the AI tool influence students’ understanding and application of requirements quality criteria?
    \item [RQ3] How do students perceive the usefulness, trustworthiness, and limitations of the AI tool, and to what extent do they critically evaluate its outputs?
\end{description}

The study uses a mixed-methods design in the RE course (N = 100; 72 analyzed submissions), combining quantitative analysis of requirements quality assessments with qualitative analysis of interaction patterns and reflective responses. By triangulating observable behavior, performance measures, and self-reported experiences, we provide empirical evidence of how structured AI integration shapes learning processes in RE.

This study contributes to SE education research in three ways. First, it applies TPACK in the context of AI-supported REE through a replicable assignment design. Second, it provides empirical evidence that structured instructional sequencing can shape students’ AI enactment toward evaluative and reflective engagement. Third, it identifies differential effects of AI support across requirements quality dimensions, offering design implications for integrating generative AI into conceptually intensive SE subjects.

By shifting attention from AI capability to pedagogical orchestration, this work advances evidence-based guidance for responsible and analytically grounded AI integration in software engineering education.

\section{Background and Related work}

Requirements engineering (RE) is a core subject in software engineering curricula, focusing on eliciting, analyzing, specifying, validating, and managing stakeholder needs. REE places strong demands on students’ abilities to reason about ambiguity, communicate with stakeholders, and work effectively with textual description. These make RE conceptually demanding and particularly sensitive to changes introduced by generative AI technologies. While AI tools can generate syntactically plausible requirements for a given software project, the development of analytical judgment and quality reasoning remains a central educational objective.

Recent advances in LLMs have influenced RE practice. A systematic literature review by Cheng et al. \cite{cheng2025generative} synthesizes the application of AI across RE activities and highlights persistent challenges related to trust and human–AI collaboration. These developments have motivated initial research on exploring AI use in REE. Guardado \textit{et al.} \cite{Guardado_RE_2025} reported that guided LLM use can improve students’ comprehension of RE practices, yet also raise concerns regarding academic integrity, over-reliance, and insufficient critical evaluation of AI outputs. Similarly, Tiwari and Rathore \cite{tiwari2025leveraging} propose structured approaches for integrating LLMs into REE, which emphasize the need for deliberate instructional design. Furthermore, practice-oriented RE research further demonstrates the increasing capability of LLMs to support RE workflows \cite{wei2024requirements}. Such studies reinforce the need to prepare students for AI-supported RE practice. Nevertheless, systematic investigations of pedagogically grounded AI integration in REE remain limited \cite{vierhauser2024towards}.

\subsection{AI in Software Engineering Education}


Beyond RE, a growing body of research examines AI integration in SE and engineering education. Sah \textit{et al.} \cite{Sah_CSEET_2024} provides a comprehensive review of AI adoption in SE education. Their findings indicate that while experimentation is widespread, many implementations lack robust pedagogical grounding. Identified challenges include instructor readiness, ethical concerns, assessment validity, and alignment between AI use and learning objectives. Similarly, Filippi and Motyl \cite{Filippi_Motyl_2024} present a growing interest in LLM applications across engineering education, while noting a lack of structured guidelines and pedagogical frameworks for effective AI integration. Studies in broader engineering education contexts, such as \cite{vidalis2024revolutionizing}, report generally positive student perceptions of AI-supported learning but stress the importance of careful instructional design.

Researchers argue that AI fundamentally challenges educational practices. Kirova et al. \cite{kirova2024software} contend that SE education must adapt assignment and assessment strategies to LLM-rich environments, while early exploratory studies \cite{levy2025work} highlight potential benefits of AI-supported learning while emphasizing the need for more systematic and theory-driven research. Overall, these findings suggest that AI adoption in SE education is advancing faster than the development of pedagogically grounded integration strategies. While AI tools demonstrate technical capability, less attention has been devoted to how instructional design mediates students’ interaction with AI-generated artifacts, particularly in conceptually intensive subjects such as RE.

\subsection{Pedagogical Framework for AI Integration and Research Gap}

The Technological Pedagogical Content Knowledge (TPACK) framework \cite{koehler2009technological} provides a well-established theoretical foundation for analyzing technology-enhanced learning. TPACK conceptualizes that effective educational use of technology arises from the integration of content knowledge (CK), pedagogical knowledge (PK), and technological knowledge (TK). 
Rather than treating technology as an isolated enhancement, TPACK emphasizes that meaningful learning emerges when technological affordances are deliberately aligned with disciplinary content and instructional strategies.

Recent extensions of TPACK to AI-enabled contexts emphasize the importance of AI literacy, trust, and ethical awareness in shaping educators’ acceptance and effective use of AI tools \cite{AI_TPACK_2024}. L. Eyal \cite{AI_Teacher_Trust_2024} emphasizes that effective AI integration requires educators to develop AI-specific technological, pedagogical, and content knowledge and to purposefully align AI use with instructional objectives. However,
existing AI-TPACK research primarily focuses on general or K-12 educational contexts. Its application in higher education, particularly in technically intensive domains, such as SE, remains underexplored.

Despite increasing experimentation with AI tools in SE education, two limitations remain evident. First, AI integration in SE
education remains largely exploratory and insufficiently grounded in pedagogical design \cite{Sah_CSEET_2024, Filippi_Motyl_2024, vierhauser2024towards}. Second, although TPACK offers a useful lens for technology integration, its application to AI-supported learning in higher education, particularly in REE involving requirements quality framework such as INVEST \cite{INVEST} and ISO/IEC/IEEE 29148 \cite{6146379} is underexplored. 

From a TPACK perspective, this gap reflects limited alignment among technological affordances (AI capability), pedagogical scaffolding (structured learning and assessment strategies), and disciplinary content knowledge (requirements quality criteria). In REE specifically, there is little empirical evidence on how AI tools can be embedded into assignments in ways that support structured analysis, reflection, and iterative refinement without undermining students’ analytical competencies. 

To address the gaps, this 
study explicitly designs and evaluates a TPACK-guided integration of a multi-agent AI tool into an RE course assignment, examining how technological, pedagogical, and content elements interact to shape students’ AI use and learning outcomes. 

\section{Research Context}
This section introduces the context and the AI tool used in the study. It provides the necessary background to motivate the subsequent research design, including the conceptual integration of the AI tool into course assignments and the associated data collection and analysis procedures.

\subsection{Course Description}

The study was conducted in the Requirements Engineering course, a 5 ECTS 
course in the Master’s Degree Programme in Computing Sciences and Electrical Engineering at Tampere University, with approximately 100 students enrolled annually. 

The course covers core RE activities in software system development. It integrates both principles and practical approaches to requirements elicitation, analysis, specification, validation, and requirements management. Students are expected to possess prior knowledge of fundamental software engineering concepts.

The course comprises 20 hours of lectures, ten weekly individual assignments, two mastery exercises, and a collaborative group work. All learning materials, including lecture notes, supplementary readings, assignment descriptions, and project guidelines, are made available through the university’s Moodle platform. Students complete individual assignments based on their understanding of lectures and textbooks. After six weeks of lectures, students start collaboratively working on a project in groups of three or four, selecting a predefined or self-proposed topic. The project allows students to apply RE knowledge in a research or practice-oriented context. In addition, a mid-term and a final mastery exercise, both given in forms of online quizzes, are used to assess students’ understanding of core RE concepts.

Although the assignments, group project, and mastery exercises are non-compulsory, active participation is essential for achieving a good final grade. This structured yet flexible course design aims to foster analytical reasoning, critical thinking, and collaboration, and prepares students to address the inherent complexity and ambiguity of real-world RE practice.

\subsection{A Multi-Agent AI Tool}

\begin{figure*}
    \centering
    \includegraphics[width=\textwidth,height=\textheight,keepaspectratio]{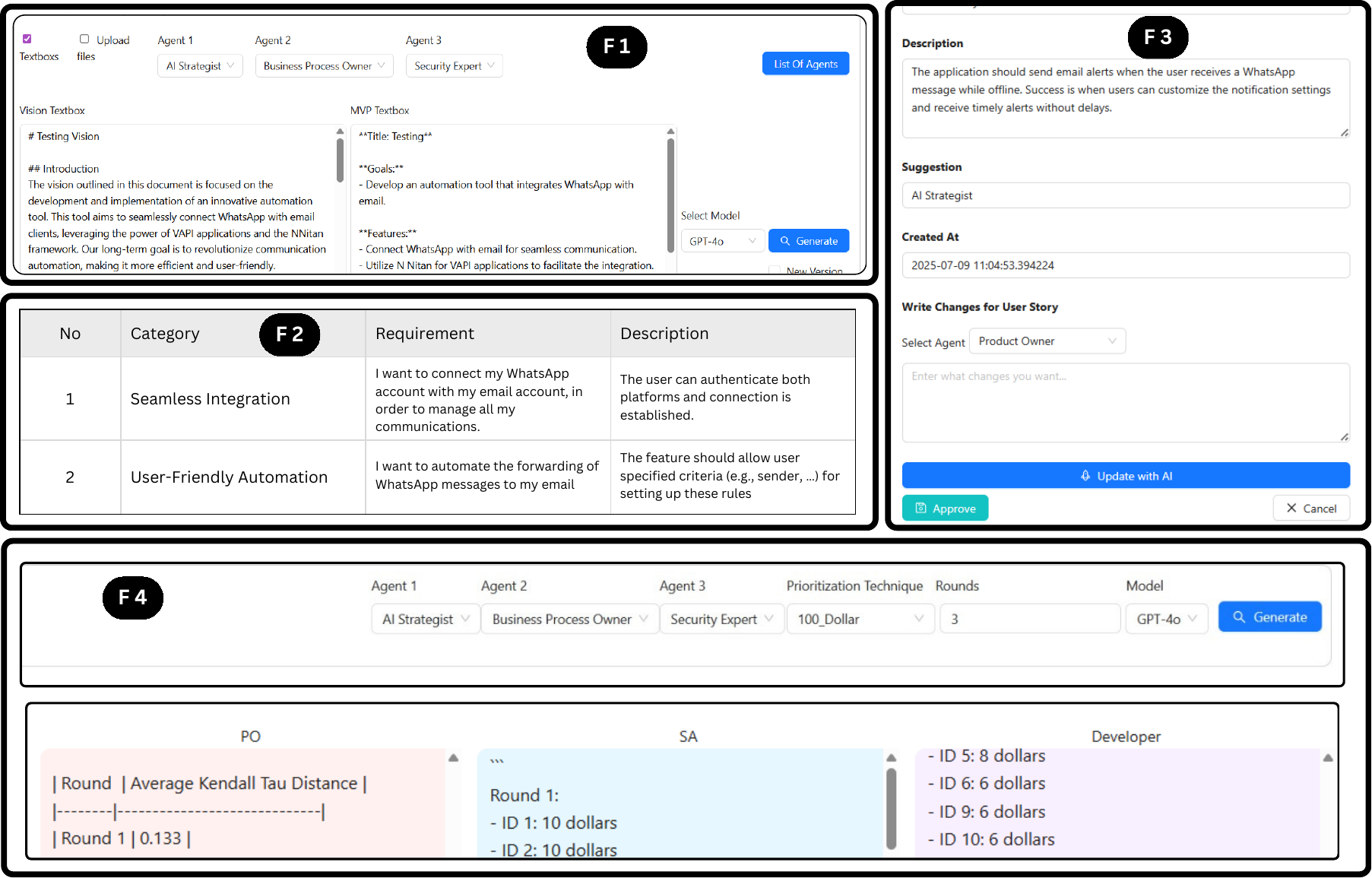}
    \caption{Screenshots of features implemented in the multi-agent AI tool}
    \label{fig:features}
\end{figure*}

To support students' learning in the REQ, the course integrated a multi-agent AI-based requirements assistance tool designed to facilitate requirements 
generation and analysis through iterative refinement. The tool \cite{sami2024early}\cite{sami2026bridging} was developed as a research prototype in the Software Engineering Research Center (TASE)\footnote{https://research.tuni.fi/tase/} at Tampere University. Its development was motivated by ongoing research on how LLMs can be orchestrated through multiple specialized agents to assist RE tasks. As illustrated in Figure \ref{fig:features}, the tool implements four features: 
\begin{enumerate}[label= \textbf{F\arabic*}, leftmargin=2em]
    \item \textbf{Agent configuration: }Users can select agents representing distinct stakeholders and specify their roles for task completion.
    \item \textbf{Requirements generation: }Based on project-specific input such as  product vision, minimal viable product description, and target users, selected agents collaboratively generate an initial set of user stories. 
    \item \textbf{Requirements analysis and refinement: }Users can iteratively refine the generated user stories through feedback cycles and manual edits. 
    \item \textbf{Requirements prioritization: }Once requirements are finalized and approved, selected agents collaboratively prioritize requirements. It is worth noting that this feature is not used in the assignment investigated in this study
\end{enumerate}


The integration of this AI tool into the course was motivated by its explicit support for requirements generation and refinement. Writing high-quality requirements is one learning objective of the course and is also an area where students commonly experience difficulties, such as formulating clear, consistent, and well-structured requirements. The tool’s multi-agent design, which exposes students to multiple stakeholder perspectives, aligns closely with the pedagogical goal of supporting students in reasoning about requirements quality, completeness, and trade-offs, rather than merely producing final student assignment submissions.

\section{Research Design}

This study adopts a mixed-methods design to investigate (i) how a multi-agent AI tool can be pedagogically integrated into an RE course assignment (RQ1), (ii) its impact on students’ understanding of requirements quality (RQ2), and (iii) students’ perceptions and evaluative engagement (RQ3). The design combines quantitative analysis of assignment submissions with qualitative analysis of student reflections, enabling triangulation between observed performance and self-reported learning experiences.

\subsection{Pedagogical Design of AI Integrating} 

The AI-integrated assignment was designed using the TPACK framework to explicitly align content knowledge (requirements quality criteria), pedagogical knowledge (contrastive and reflective learning activities), and technological knowledge (AI-based requirements generation and refinement). The design aimed to ensure that AI use supported structured analysis and iterative improvement rather than automation.


The AI tool was integrated into 
an individual assignment focusing on reviewing and improving requirements. Students reviewed requirements elicited in earlier assignments for the same project context and aligned them with quality criteria such as INVEST framework \cite{INVEST} and related guidelines.

This assignment was selected for AI integration because prior course offerings revealed recurring learning challenges. Despite explaining the concepts with examples through lectures, students tended to specify requirements that were vague, solution-oriented, insufficiently testable, or lacking explicit justification of value, and struggled to systematically assess and improve the requirements.

The AI tool was introduced only after manual revision, allowing students to generate alternative user stories for the same functional goals and compare them with their own revisions. This contrastive design encouraged analytical evaluation and iterative refinement. The expected workload was approximately 2–3 hours.


\subsubsection{Task Workflow}

\begin{figure}[h]
    \centering
    \includegraphics[width=\columnwidth]{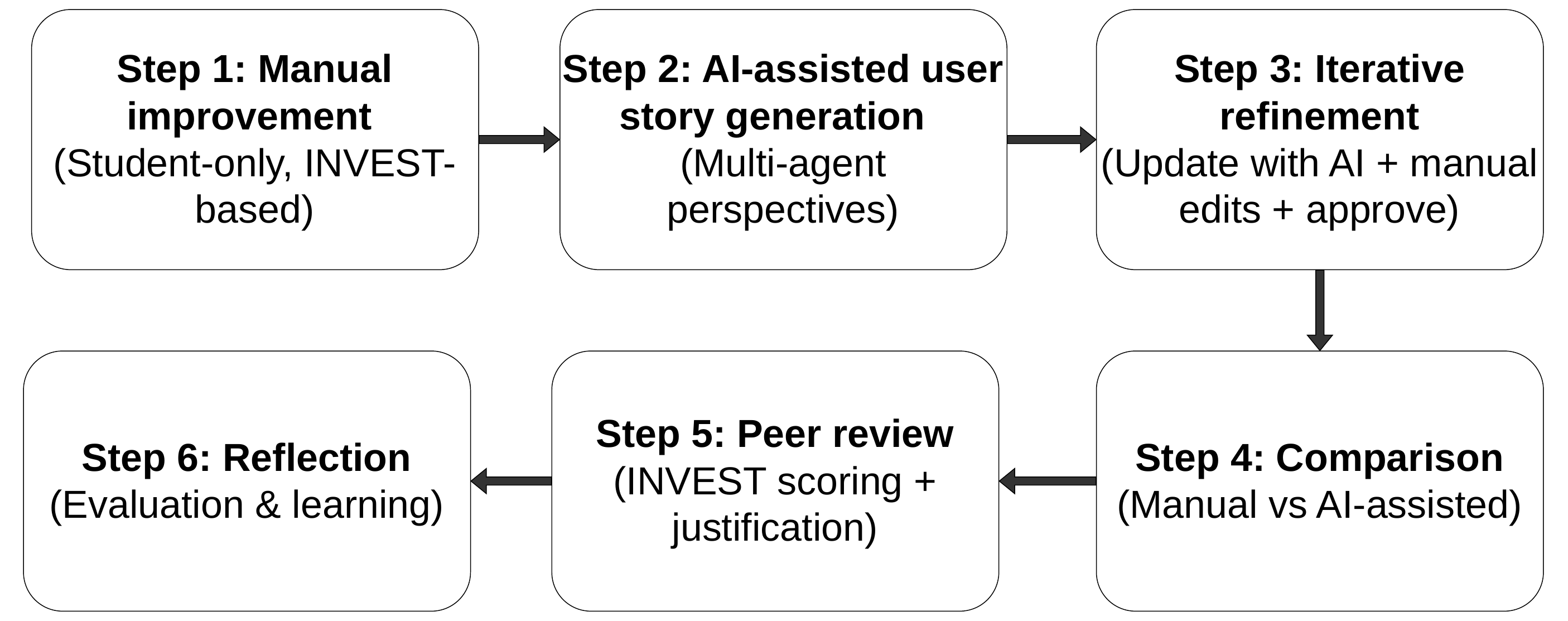}
    \caption{Assignment workflow for AI tool integration in Assignment~4}
    \label{fig:assignment_workflow}
\end{figure}

The assignment followed a staged workflow designed to support contrastive learning and to discourage automation-oriented use of the AI tool. As illustrated in Figure \ref{fig:assignment_workflow}, each stage required students to actively engage in evaluation, comparison, and refinement of requirements rather than relying on AI-generated outputs.

\begin{enumerate}[label=Step \arabic*, leftmargin=3em]
    \item \textbf{Manual improvement:} Students 
    selected a subset of 
    requirements from earlier assignments and manually revised them as user stories, following the INVEST quality framework that includes quality attributes of \emph{Independent}, \emph{Negotiable}, \emph{Valuable}, \emph{Estimable}, \emph{Small}, and \emph{Testable} \cite{INVEST}.
    \item \textbf{AI-assisted user story generation:} Students used the AI tool to generate user stories by configuring persona agents representing different stakeholders and providing project-specific inputs, including the product vision, user groups, and other relevant descriptions documented in prior assignments. Tool instructions\footnote{https://doi.org/10.6084/m9.figshare.31440988} were provided on the course Moodle page. The instructions were intended to familiarize students with the tool while leaving configuration choices to the students’ discretion.
    \item \textbf{Iterative refinement:} Students reviewed and refined AI-generated user stories through feedback cycles, manual edits, addition or removal of stories, and explicit approval actions.
    \item \textbf{Comparison:} Students compared manually revised and AI-generated user stories, focusing on differences in clarity, specificity, and alignment with INVEST attributes. The user stories were documented in a shared spreadsheet\footnote{https://doi.org/10.6084/m9.figshare.31430224}, which is used for the subsequent peer-review step. 
    \item \textbf{Peer review:} Students evaluated their peer student’s requirements documented on the shared spreadsheet, using structured INVEST-based scoring with justifications. The authors of the reviewed stories were expected to respond with agreement or disagreement.
    \item \textbf{Reflection:} After completing the peer-review step,  students completed a reflection questionnaire\footnote{https://doi.org/10.6084/m9.figshare.31430221} to document their interactions with the tool and their perceived strengths and limitations of the AI tool.
\end{enumerate}


In addition to the main assignment workflow, a pre-assignment and a follow-up mastery exercise were deliberately integrated. In both tasks, students analyzed four user stories (US1–US4) to identify quality violations against the six dimensions of the INVEST framework. To provide a benchmark, the instructor and teaching assistants together established a reference analysis, i.e. evaluation of each user story across the INVEST dimensions\footnote{https://doi.org/10.6084/m9.figshare.31442422}. The repeated tasks, compared against the predefined reference values, allowed us to compare students’ analytical reasoning before and after the AI-integrated assignment workflow.




\subsubsection{Roles of the AI tool and Students}

Within this design, the AI tool assumed the role of a supporting assistant or alternative analyst, comparable to a stakeholder or domain expert capable of proposing candidate requirements based on provided context and feedback. The tool generated alternatives and responded to refinement prompts but did not make final decisions.

Students, in contrast, assumed multiple active roles aligned with the learning objectives of the assignment. They acted as product owners when manually articulating requirements, as 
a project manager when engaging multiple stakeholder perspectives through agent selection, as requirements analysts when refining and approving AI-generated outputs, and as reviewers during comparison and peer-review phases. This role-based design positioned students as responsible evaluators and decision-makers, ensuring that learning outcomes related to structured analysis, reflection, and iterative refinement remained firmly under student control.
    
\subsection{Data Collection}

The RE course was delivered from September 2 to December 1, 2025, with a total enrollment of 100 students. The designed assignment was released on September 24 as Assignment 4 within a sequence of ten individual assignments and remained open for a two-week completion period. Data were primarily collected from \textit{student assignment submissions} in Steps 1-5 and the \textit{reflection questionnaire} completed in Step 6. In addition, the data on students' quality assessment of US1-US4 were collected in the pre-assignment and the mastery exercise. 


\textit{Student assignment submissions} were collected using a shared spreadsheet. The collected data included manually revised requirements, AI tool generated and subsequently refined user stories, peer-review quality scores, and students’ responses indicating agreement or disagreement with peer feedback. These data provide observable traces of student interaction with the AI tool, including selection, refinement, evaluation, and justification behaviors. 

The \textit{reflection questionnaire} was designed to collect students’ perspectives on their interaction with the AI tool. It served as complementary data that are not directly observable from the assignment submissions alone, such as reasoning behind refinement decisions, criteria used for acceptance or rejection of AI outputs, and perceived learning benefits or limitations. 
The questionnaire addressed: (i) students’ interaction patterns with the AI tool, e.g., time spent, number of generated and approved user stories, refinement actions, etc.; (ii) students’ choices and rationale for configuring AI agents in different stakeholder roles; (iii) perceived impact of the tool on students’ understanding and application of requirements quality attributes; (iv) students’ evaluation of AI-generated outputs, including editing, acceptance, or rejection decisions; and (v) perceptions of usability, usefulness, confidence, and limitations of the AI tool. In total, the questionnaire comprised 20 mandatory close-ended items and five optional open-ended questions.
 


\subsection{Data Analysis}

Descriptive statistics and qualitative thematic coding were applied to examine learning impact and student perceptions. Student interaction with the AI tool was analyzed using metrics such as generation counts, approval rates, and refinement frequencies to characterize engagement patterns.

To evaluate changes in requirements quality reasoning, students’ INVEST-based assessments of US1–US4 were compared before and after the AI-integrated assignment. Agreement proportions relative to predefined reference analyses were calculated, and changes in alignment ($\Delta = p_{after} - p_{pre}$) were examined for each INVEST quality dimension.

Questionnaire data were analyzed using distributional summaries of Likert-scale responses. Open-ended responses were coded thematically following established approaches to thematic analysis \cite{BraunClarke2006} to identify patterns related to perceived usefulness, evaluative behaviors, trust, and reported limitations. Cross-tabulation analyses explored relationships between reported refinement behaviors and perceived improvements across quality dimensions. This triangulated approach \cite{CreswellPlanoClark2018} enabled an integrated interpretation of pedagogical design, observed interaction patterns, learning outcomes, and student perceptions. 

\section{Results}

Although the assignment was non-compulsory, 74 out of 100 students completed it, of whom 72 provided informed consent and were included in the analysis. The average completion time for the reflection questionnaire was 27 minutes and 45 seconds. This participation level indicates substantial voluntary engagement with both the assignment and reflection components, providing a sufficiently rich dataset to examine the pedagogical integration of the AI tool and students’ enactment of the intended assignment design in practice.

\subsection{RQ1 - Students' Engagement and Interaction Patterns}


The staged assignment workflow reflects an alignment of technological, pedagogical, and content elements as conceptualized in the TPACK framework. The assignment design includes several explicit pedagogical mechanisms that guide students’ engagement with the tool. These included (i) a sequenced task structure requiring manual requirement improvement prior to AI use, (ii) explicit grounding in the INVEST quality framework, (iii) an iterative refinement loop with explicit approval actions, (iv) multi-agent personas representing diverse stakeholder perspectives, and (v) mandatory peer review requiring justification of agreement or disagreement.

These mechanisms were aligned with three intended learning processes. Structured analysis was supported through requirements quality evaluation criteria and peer-review. Reflection was embedded via comparison tasks, disagreement justification, and post-task reflection prompts. Iterative refinement was encouraged through repeated AI-assisted updates and selective approval requirements. Together, these defined the intended pedagogical role of the AI tool as a support for analytical reasoning and evaluative judgment rather than as an automation mechanism.


Table~\ref{tab:interaction_summary} summarizes key interaction metrics describing students' engagement with the AI tool during the assignment. The tool generated a median of 16.5 user stories, of which a median of 9.5 stories were approved by students, corresponding to a median approval rate of 56\%. This pattern indicates selective filtering rather than general acceptance of AI-generated outputs. Students completed a median of 1.5 refinement rounds, indicating iterative engagement with the generated requirements. Additionally, 24\% of students (n = 17) explicitly reported editing AI outputs, providing further evidence of active refinement behavior.

\begin{table}[h]
\centering
\caption{Summary of Student Interaction with the AI Tool (N=72)}
\label{tab:interaction_summary}
\begin{tabular}{l c}
\hline
\textbf{Metric} & \textbf{Value} \\
\hline
Median AI-generated stories per student & 16.5 \\
Median AI-approved stories per student & 9.5 \\
Median approval rate & 56\% \\
Median refinement rounds & 1.5 \\
Students who edited AI outputs & 24\% \\
\hline
\end{tabular}
\end{table}

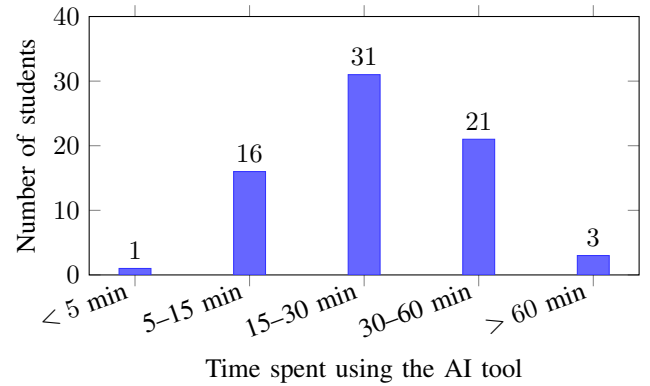
\begin{figure}[h]
\centering
\begin{tikzpicture}
\begin{axis}[
    ybar,
    bar width=12pt,
    ymin=0,
    ymax=40, 
    ylabel={Number of students},
    xlabel={Time spent using the AI tool},
    symbolic x coords={lt5,m5_15,m15_30,m30_60,gt60},
    xtick=data,
    xticklabels={$<5$ min, 5--15 min, 15--30 min, 30--60 min, $>60$ min},
    xticklabel style={rotate=20, anchor=east},
    nodes near coords,
    nodes near coords align={vertical},
    width=\columnwidth,
    height=5cm
]
\addplot[
    fill=blue!60,
    draw=blue!80
] coordinates {
    (lt5,1)
    (m5_15,16)
    (m15_30,31)
    (m30_60,21)
    (gt60,3)
};
\end{axis}
\end{tikzpicture}
\caption{Distribution of time spent using the AI tool during Assignment~4 (N=72).}
\label{fig:time_spent_distribution}
\end{figure}

Figure~\ref{fig:time_spent_distribution} illustrates the distribution of time spent using the tool. 52 students (72\%) reported 15–60 minutes of use, aligning with the expected workload. A smaller group of 17 students (24\%) spent less than 15 minutes, suggesting more surface-level interaction, wheras 3 students spent more than 60 minutes, often associated with experimentation or troubleshooting described in the reflection responses.

\begin{figure}[h]
\centering
\begin{tikzpicture}
\begin{axis}[
    ybar,
    bar width=12pt,
    ymin=0,
    ymax=35, 
    ylabel={Number of students},
    xlabel={Number of refinements made},
    symbolic x coords={zero,one_two,three_five,gt_five},
    xtick=data,
    xticklabels={0, 1--2, 3--5, $>5$},
    nodes near coords,
    nodes near coords align={vertical},
    width=\columnwidth,
    height=5cm
]
\addplot[
    fill=blue!60,
    draw=blue!80
] coordinates {
    (zero,18)
    (one_two,29)
    (three_five,19)
    (gt_five,6)
};
\end{axis}
\end{tikzpicture}
\caption{Distribution of the number of refinements made by students during Assignment~4 (N=72).}
\label{fig:refinement_distribution}
\end{figure}
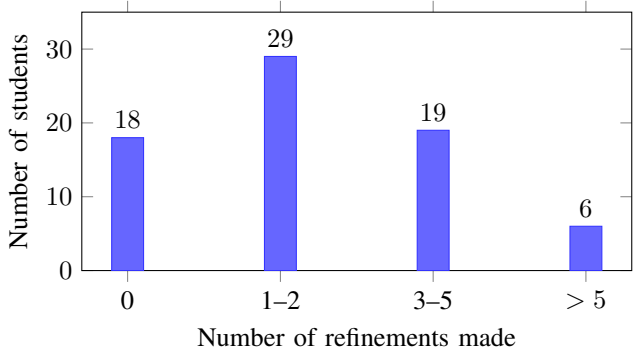

Figure~\ref{fig:refinement_distribution} presents the refinement frequency. Overall, 75\% of students modified at least one AI-generated user story, with 29 students performing 1–2 refinement rounds and 25 students performing 3 or more. Only 18 students made no refinements. Together with the 56\% approval rate of the AI tool generated user stories, these results indicatethat most students engaged in evaluative selection and modification rather than passive adoption.

\subsection{RQ2 - Learning Impact on Requirements Quality Assessment}

Learning impact was assessed by comparing students’ evaluations of four user stories (US1–US4) before and after the AI-integrated assignment. For each INVEST dimension, agreement with instructor-defined reference values was calculated, and alignment changes ($\Delta = p_{after} - p_{pre}$) were computed.

Table~\ref{tab:rq2_numeric_improvement} shows that alignment changes are quality dimension- and story-specific rather than uniformly positive. The most consistent improvements appear for \emph{Valuable}, which is positive in US2 to US4, and \emph{Testable}, which is positive in US1 to US3, including a notable increase of 0.108 in US1. On the other hand, \emph{Small}, \emph{Independent}, and \emph{Estimable} exhibit mixed patterns across stories. 
Negotiable shows non-positive change across all four stories, with the largest decrease in US2 at -0.292. 

\begin{table}[h]
\centering
\caption{Agreement change in INVEST evaluation ($\Delta = p_{after} - p_{pre}$)}

\label{tab:rq2_numeric_improvement}
\begin{tabular}{lcccc}
\hline
\textbf{INVEST dimension} & \textbf{$\Delta_{US1}$} & \textbf{$\Delta_{US2}$} & \textbf{$\Delta_{US3}$} & \textbf{$\Delta_{US4}$}  \\
\hline
Independent & \underline{0.000} & -0.153 & \textbf{0.194} & -0.014  \\
Negotiable & \underline{0.000} & -0.292 & -0.042 & -0.127  \\
Valuable   & -0.041 & \textbf{0.014} & \textbf{0.014} & \textbf{0.042}  \\
Estimable  & -0.081 & \textbf{0.069} & \textbf{0.028} & -0.113  \\
Small      & -0.014 & \textbf{0.125} & \textbf{0.028} & -0.155  \\
Testable   & \textbf{0.108} & \textbf{0.014} & \textbf{0.028} & -0.014 \\
\hline
\end{tabular}
\end{table}

The contrast between US3 and US4 provides additional insight. US3, which violated multiple INVEST dimensions in the reference analysis, shows positive alignment changes across most dimensions, indicating increased convergence toward the reference evaluation. In contrast, US4, which did not violate any dimension in the reference evaluation, exhibits limited improvement and slight divergence in some dimensions. Given the absence of clear flaws in US4, these shifts reflect greater variability in students' evaluations.



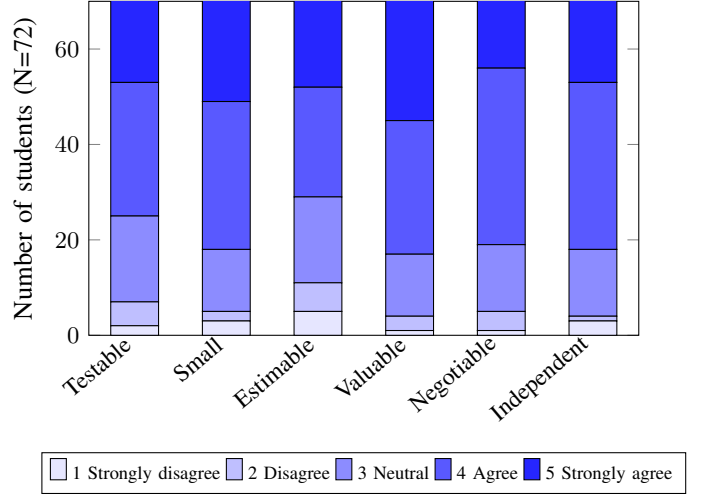
\begin{figure}[ht]
\centering
\begin{tikzpicture}
\begin{axis}[
    width=\columnwidth,
    height=6cm,
    ybar stacked,
    bar width=18pt,              
    ymin=0, ymax=70,
    ylabel={Number of students (N=72)},
    symbolic x coords={
        Testable,
        Small,
        Estimable,
        Valuable,
        Negotiable,
        Independent
    },
    xtick=data,
    xticklabel style={
        rotate=40,              
        anchor=east,
        font=\small
    },
    enlarge x limits=0.10,      
    tick label style={font=\small},
    legend style={
        at={(0.5,-0.35)},       
        anchor=north,
        legend columns=5,
        font=\scriptsize,
        draw=black
    },
]

\addplot+[fill=blue!10, draw=black] coordinates {
    (Testable,2) (Small,3) (Estimable,5)
    (Valuable,1) (Negotiable,1) (Independent,3)
};
\addplot+[fill=blue!25, draw=black] coordinates {
    (Testable,5) (Small,2) (Estimable,6)
    (Valuable,3) (Negotiable,4) (Independent,1)
};
\addplot+[fill=blue!45, draw=black] coordinates {
    (Testable,18) (Small,13) (Estimable,18)
    (Valuable,13) (Negotiable,14) (Independent,14)
};
\addplot+[fill=blue!65, draw=black] coordinates {
    (Testable,28) (Small,31) (Estimable,23)
    (Valuable,28) (Negotiable,37) (Independent,35)
};
\addplot+[fill=blue!85, draw=black] coordinates {
    (Testable,19) (Small,23) (Estimable,20)
    (Valuable,27) (Negotiable,16) (Independent,19)
};

\legend{
1 Strongly disagree,
2 Disagree,
3 Neutral,
4 Agree,
5 Strongly agree
}

\end{axis}
\end{tikzpicture}
\caption{Distribution of student Likert-scale responses for perceived support across INVEST dimensions.}
\label{fig:rq3_likert_invest}
\end{figure}

Figure~\ref{fig:rq3_likert_invest} complements these findings by presenting students’ perceived AI tool support of understanding across the INVEST dimensions. The Likert distributions are predominantly positive, with roughly two-thirds to three-quarters of students agreeing that the AI tool supported improvements in most dimensions, including \emph{Negotiable}. 

A divergence arises between the measured alignment and the perceived improvement. While \emph{Valuable} and \emph{Testable} display both positive alignment shifts and strong perceived support, \emph{Negotiable} shows high perceived support despite declining alignment with the reference evaluation. This pattern indicates that perceived improvement does not necessarily correspond to convergence with the instructor-defined interpretation of specific quality attributes.

\subsection{RQ3 - Student Perceptions: Usefulness, Trust, Critical Thinking}

Building on the learning effects reported in RQ2, we next examine how students perceived the AI tool and the extent to which they critically evaluated its outputs,  using Likert-scale responses and thematic coding of open-ended reflections.

\begin{figure}[ht]
\centering
\begin{tikzpicture}
\begin{axis}[
    width=\columnwidth,
    height=6cm,
    ybar stacked,
    ymin=0, ymax=70,
    bar width=24pt,
    ylabel={Number of students (N=72)},
    symbolic x coords={
        {Ease of\\use},
        {Critical\\thinking},
        {Confidence in\\AI outputs}
    },
    xtick=data,
    xticklabel style={
        align=center,
        font=\small
    },
    enlarge x limits=0.25,
    legend style={
        at={(0.5,-0.28)},
        anchor=north,
        legend columns=5,
        font=\scriptsize,
        draw=black
    },
]

\addplot+[fill=blue!10, draw=black] coordinates {
    ({Ease of\\use},6)
    ({Critical\\thinking},4)
    ({Confidence in\\AI outputs},4)
};
\addplot+[fill=blue!25, draw=black] coordinates {
    ({Ease of\\use},10)
    ({Critical\\thinking},6)
    ({Confidence in\\AI outputs},4)
};
\addplot+[fill=blue!45, draw=black] coordinates {
    ({Ease of\\use},20)
    ({Critical\\thinking},8)
    ({Confidence in\\AI outputs},23)
};
\addplot+[fill=blue!65, draw=black] coordinates {
    ({Ease of\\use},25)
    ({Critical\\thinking},21)
    ({Confidence in\\AI outputs},17)
};
\addplot+[fill=blue!85, draw=black] coordinates {
    ({Ease of\\use},11)
    ({Critical\\thinking},33)
    ({Confidence in\\AI outputs},24)
};

\legend{
1 Strongly disagree,
2 Disagree,
3 Neutral,
4 Agree,
5 Strongly agree
}

\end{axis}
\end{tikzpicture}
\caption{Distribution of student Likert-scale responses for perceived usability, critical-thinking support, and confidence in AI outputs).}
\label{fig:rq3_likert_noninvest}
\end{figure}
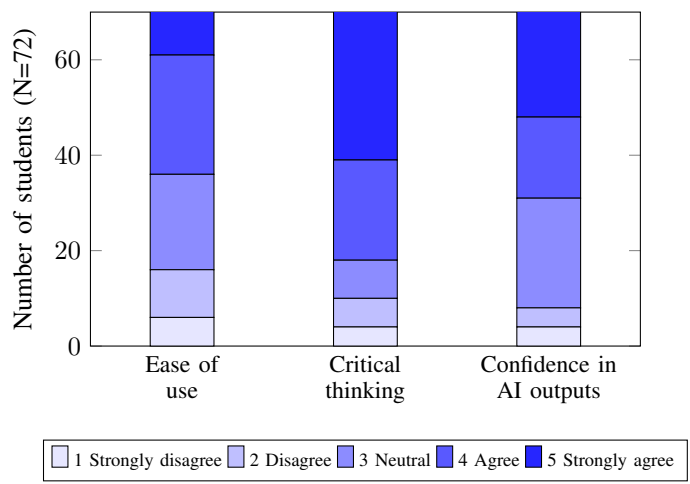

Figure~\ref{fig:rq3_likert_noninvest} presents distributions for students' perceived usability, critical-thinking support, and confidence in AI outputs. Perceived usability was moderately positive, with 36 students selecting agreement categories and 16 expressing negative perceptions, indicating variability in user experience. Perceived support for critical thinking received the highest ratings, with 52 students selecting agreement categories. Confidence in AI-refined outputs was also generally positive, with 40 agreement responses. These quantitative patterns suggest that students largely viewed the tool as cognitively supportive, though not uniformly smooth in interaction.

Table~\ref{tab:rq3_qualitative_summary} summarizes themes identified from open-ended responses, with each theme counted once per student.

\begin{table}[h]
\centering
\caption{Reflection themes (N=72)}
\label{tab:rq3_qualitative_summary}
\begin{tabular}{p{6.5cm} c}
\hline
\textbf{Construct and Representative Aspects} & \textbf{Students (n)} \\
\hline

\textbf{Usefulness} & \\
\quad Idea generation / helpful suggestions & 47 \\
\quad Added detail / increased specificity & 32 \\
\quad Increased INVEST awareness & 30 \\
\quad Acceptance-criteria support & 13 \\

\\
\textbf{Confidence in AI outputs} & \\
\quad Explicit distrust toward AI outputs & 4 \\
\quad Active verification against INVEST & 14 \\
\quad Explicitly challenged or rejected outputs & 10 \\

\\
\textbf{Critical evaluation and refinement} & \\
\quad Edited for improved testability & 21 \\
\quad Edited for improved value alignment & 17 \\
\quad Edited for smaller, manageable stories & 13 \\
\quad Edited for independence or negotiability & 16 \\

\\
\textbf{Perceived limitations} & \\
\quad Usability / interface issues & 22 \\
\quad Context mismatch or vague outputs & 13 \\
\quad Overly large or duplicated requirements & 8 \\

\hline
\end{tabular}
\end{table}

\textbf{Usefulness} was most frequently associated with idea generation and increased specificity. 47 students described the tool as helpful for generating or structuring ideas, 32 reported increased detail in their requirements, and 30 indicated heightened awareness of missing INVEST attributes. Thirteen explicitly mentioned support for defining acceptance criteria. One student summarized this perceived benefit: \textit {“The AI-generated stories provided structured language, clearer acceptance criteria, and generally better adherence to the INVEST framework”}. These perceptions align with the high Likert ratings for improvements in \emph{Valuable}, \emph{Independent}, and \emph{Small} dimensions.

\textbf{Confidence in AI outputs} was generally positive but conditional. Explicit distrust was rare (n = 4). However, trust was not unconditional: 14 students reported actively verifying outputs against the INVEST framework, and 10 explicitly rejected or questioned AI-generated requirements. As one student noted: \textit {“Occasionally, the AI-generated stories included generic language that was not specific to drone delivery ... I had to manually refine these outputs to make them fully relevant to the project scenario”}. Such responses indicate cautious reliance, where AI suggestions were treated as drafts requiring further validation.

\textbf{Critical thinking and evaluation} was demonstrated by evidence of active refinement of requirements generated by the tool. 21 students reported modifying outputs to improve testability, 17 strengthened value alignment, and 13 reduced overly broad requirements. These behaviors correspond with the strong Likert ratings for critical-thinking support (52 agreement responses). Cross-tabulation further indicates alignment between editing and perceived improvement; for example, 15 of 21 students who edited for testability (71\%) agreed that the tool improved testability, compared to 47 of 72 students (65\%) in the overall sample. One reflection illustrates this process: \textit {My edits were mainly focused on making the stories more \emph{Testable} and \emph{Valuable} by adding measurable success criteria ... I also improved Independence by narrowing broad stories ... and removed vague terms like “useful” or “engaging”.}

\textbf{Usability} and contextual mismatch were also reported. 22 students reported interface or workflow issues, and 13 described vague or contextually inappropriate outputs. 8 reported overly broad or duplicated requirements. Despite these limitations, distrust remained limited, as one student observed: \textit {"The quality of answers was generally okay, but the tool tended for some reason to favour the Product owner ... I had to run the tool several times (after the first attempt) before developer requirements started appearing"}. Such statements indicate frustration with tool behavior rather than rejection of its conceptual value.

Overall, students perceived the AI tool primarily as a scaffold for structured reflection and quality-oriented refinement rather than as an automated solution generator. Their responses indicate conditional trust, active verification, and dimension-specific engagement with requirements quality criteria, even when usability challenges were present.


\section{Discussion}
\
This study examined how a multi-agent AI tool can be pedagogically integrated into a REQ course, through a TPACK-guided pedagogical design, and how such integration shapes student interaction, learning outcomes, and evaluative engagement.

\subsection{Pedagogical Design Shapes AI Use}

The study shows that students’ AI use was shaped primarily by instructional design rather than by technological capability alone. Although the multi-agent AI tool enabled requirements generation and refinement, students demonstrated selective approval (median approval rate: 56\%), iterative revisions, and explicit verification against the INVEST framework. These indicate evaluative filtering rather than automation-oriented acceptance.

From a TPACK perspective \cite{koehler2009technological}, technological affordances were enacted within a pedagogically structured workflow, i.e. manual-first revision, approval, peer review, and reflection, and anchored in the requirements quality framework. This alignment of technological, pedagogical, and content knowledge preserved student responsibility for quality judgment and positioned AI-generated requirements as artifacts for analysis rather than final solutions.




Where prior research has observed that AI integration in SE education often remains exploratory or insufficiently structured \cite{Guardado_RE_2025}\cite{Sah_CSEET_2024}, the present findings provide empirical evidence that deliberate pedagogical orchestration can meaningfully shape AI enactment. Rather than evaluating the AI tool in isolation, this study demonstrates how workflow design mediates the relationship between AI capability and student behavior. In the context of REE where analytical reasoning about requirements quality is essential, these results imply that effective AI integration depends less on tool sophistication than on instructional mechanisms that structure comparison, justification, and iterative refinement.


\subsection{Learning Impact}
The alignment changes in requirements quality assessment reveal differentiated learning effects across INVEST dimensions rather than uniform improvement. Notably, the results reveal a distinction between structurally explicit quality attributes such as \emph{Testable, Valuable} and interpretive attributes such as \emph{ Negotiable}. Structurally explicit attributes lend themselves to clearer textual articulation and measurable criteria, whereas interpretive attributes require contextual reasoning about stakeholder flexibility and solution openness. In particular, the divergence observed for \emph{Negotiable} suggests that students may have developed increased awareness of the attribute without fully converging toward the instructor-defined reference analysis. This divergence does not necessarily indicate regression. Instead, it may reflect increased attention to interpretive ambiguity, where students move from implicit assumptions toward more varied evaluative interpretations. In AI-supported contexts, plausible textual formulations may amplify awareness of quality dimensions without ensuring convergence toward a single normative interpretation. These findings indicate that AI tools may more readily support structurally explicit quality attributes than interpretive ones requiring contextual judgment. From a TPACK perspective, this emphasizes that technological affordances can stimulate reflection, but sustained alignment with the requirements quality criteria depends on targeted pedagogical mediation.

These findings are consistent with prior research showing that LLMs are effective at enhancing textual structure and clarity, yet less consistent in supporting nuanced conceptual reasoning and context-sensitive judgment \cite{Filippi_Motyl_2024}. In educational contexts, LLM-generated outputs often improve articulation and organization but do not automatically ensure convergence with deeper evaluative criteria without instructional scaffolding.

For practice, this suggests that interpretive quality dimensions, such as negotiability or independence, require explicit exemplars, structured comparison tasks, and guided evaluative criteria. Future research should examine how tool configuration and scaffolding strategies can better support alignment across complex quality attributes in REE.

\subsection{Student Perceptions and Conditional Trust}
Students' perceptions of the AI tool reveal the usefulness of the tool and conditional trust. While students generally reported positive perceptions, particularly for idea generation, structuring requirements, and increasing awareness of quality criteria, qualitative evidence revealed conditional trust: students verified AI outputs against INVEST criteria, edited unclear suggestions, and rejected contextually inappropriate requirements. AI-generated content was thus treated as provisional drafts rather than authoritative solutions.

The discrepancy between perceived improvement and measured alignment further sharpens this interpretation. Although students frequently reported that the tool enhanced their understanding of quality attributes, convergence with instructor-defined evaluative standards was uneven. This suggests that AI integration may initially enhance awareness and reflective attention before producing consistent normative convergence. Rather than demonstrating blind reliance, students appeared to treat AI suggestions as drafts requiring validation. This indicates that structured integration can promote critical engagement rather than over-reliance.

From a pedagogical standpoint, these findings highlight that AI can function as a scaffold for reflective practice when integrated into workflows that require explicit verification, justification, and iterative refinement. Rather than displacing analytical reasoning, structured AI integration can reinforce it. Future research should examine whether such guided engagement produces sustained development of analytical competence and judgment in RE contexts beyond short-term assignment performance.

\subsection{Implications for RE and SE Education}
The findings collectively suggest that effective AI integration in RE and SE education requires deliberate orchestration rather than permissive tool adoption. This study suggests several design principles for AI integration in RE and SE education:

\begin{itemize}
    \item \textbf{Sequence human reasoning before AI use} Manual-first workflows preserve ownership of analysis and reduce over-reliance and automation-oriented behavior.
    \item \textbf{Structure AI use around contrastive analysis} AI-generated alternatives are most effective when used as stimuli for comparison rather than as final outputs.
    \item \textbf{Anchor AI use in explicit quality frameworks} Standards such as INVEST provide evaluative reference points for assessing AI outputs.
    \item \textbf{Incorporate peer review and reflection} Peer review and reflection promote active evaluation and mitigate over-reliance.
\end{itemize}

Importantly, these principles indicate that AI integration should be differentiated according to the epistemic nature of the target learning objectives. For structurally explicit competencies, AI can serve as a productive accelerator. For interpretive and context-sensitive competencies, stronger scaffolding and explicit exemplars remain necessary.


\subsection{Threats to Validity} 
We structure potential threats following the validity framework \cite{wohlin2012experimentation}, considering internal, construct, conclusion, and external validity.

Internal validity concerns whether observed effects can be attributed to the AI-integrated assignment rather than alternative factors. The absence of a control group limits causal attribution, as improvements between the pre-assignment and mastery exercise may partly reflect course progression or repeated exposure to the INVEST framework in lectures and other assignments. Furthermore, students completed assignments independently, and full adherence to the prescribed assignment workflow could not be verified. Triangulation across assignment submissions, interaction metrics, and reflective responses partially mitigates these concerns.

Construct validity may be affected by operationalizing learning impact as agreement with instructor-defined reference analyses. While this provides a consistent benchmark, it may not capture alternative defensible interpretations or deeper conceptual understanding, particularly for interpretive dimensions such as \emph{Negotiable}. Additionally, students may emphasize different quality aspects in their evaluations, and the use of only four user stories limits the breadth of assessment. Likert-scale responses capture perceived support rather than objective learning gains; combining the questionnaire data with assignment-based evaluation and thematic coding of open-ended reflections helps mitigate this limitation.

Conclusion validity is constrained by reliance on descriptive statistics and agreement proportions, which support exploratory interpretation but limit strong causal inference. Moreover, the AI tool was developed within the same research environment as the course, potentially introducing contextual bias despite predefined evaluation criteria and structured coding procedures.

External validity is limited by the study’s conduct within a single RE course using a research prototype. Findings may not generalize to other institutional contexts, undergraduate cohorts, commercial AI systems, or team-based settings. Nonetheless, the course structure reflects common REE practices, and the high voluntary participation rate and multi-source data provide a meaningful basis for interpretation within similar contexts.

\section{Conclusion}

Integrating multi-agent AI tools into REE presents both opportunities and pedagogical challenges. This study examined a TPACK guided integration of a multi-agent AI tool into an RE course, focusing on how assignment design shaped students’ AI enactment and requirements quality reasoning.  

The findings demonstrate that AI can support students’ analytical engagement when it is embedded within a carefully designed pedagogical structure. Students did not treat the tool as an automated generator but instead selectively approved, refined, and evaluated AI-generated requirements. 

These results suggest that the effective AI integration depends less on a tool's technical sophistication and more on how it is incorporated into assignment design. Sequencing manual work before AI interaction, embedding structured comparison and peer review, and anchoring evaluation in formal quality frameworks appear to be key factors in promoting critical thinking rather than passive reliance on automation. However, usability limitations and interpretive ambiguity in certain quality dimensions highlight areas that require further instructional refinement.

Future research should investigate the long-term development of analytical competence in AI-integrated RE environments and explore how orchestration strategies can be adapted across diverse institutional contexts and AI configurations. By refining pedagogically grounded integration strategies, educators can leverage AI as a scaffold for reflective practice while preserving the analytical rigor central to SE education.

\bibliographystyle{IEEEtran} 
\bibliography{references}

\balance

\end{document}